\documentclass[twocolumn]{emulateapj}

\usepackage[varg]{txfonts}

\usepackage{times}
\usepackage{graphicx,bm,amssymb}
\usepackage{epsfig}
\usepackage{amssymb}
\usepackage{natbib}

\usepackage{psfrag}

\def\araa{ARA\&A}
\def\apj{ApJ}
\def\apjl{ApJ}
\def\apss{Ap\&SS}
\def\aap{A\&A}
\def\mnras{MNRAS}
\def\na{New A}
\def\nat{Nature}
\def\bain{Bull.~Astron.~Inst.~Netherlands}

\newcommand{\msun}{\mbox{$M_\odot$}}
\newcommand{\rsun}{\mbox{$R_\odot$}}

\def\be{\begin{eqnarray}}
\def\ee{\end{eqnarray}}
\def\bi{\begin{itemize}}
\def\ei{\end{itemize}}
\def\lsim{\mathrel{\rlap{\lower3pt\hbox{\hskip1pt$\sim$}}
     \raise1pt\hbox{$<$}}} 
\def\gsim{\mathrel{\rlap{\lower3pt\hbox{\hskip1pt$\sim$}}
     \raise1pt\hbox{$>$}}} 

\shorttitle{Weakly-Magnetized, Slowly-Rotating Progenitors of GRBs}
\shortauthors{Enrique Moreno M\'endez}

\begin{document}

\title{An Argument for Weakly-Magnetized, Slowly-Rotating Progenitors of \\Long Gamma-Ray Bursts}

\author{Enrique Moreno M\'endez}
\affil{Instituto de Astronom\'ia, Universidad Nacional Aut\'onoma de M\'exico, Circuito Exterior, Ciudad Universitaria, \\Apartado Postal 70-543, 04510, Distrito Federal, M\'exico.
\\Argelander-Institut f\"ur Astronomie, Bonn University, Auf dem H\"ugel 71, 53121 Bonn, Germany.}

\email{EMM: enriquemm@astro.unam.mx}


\begin{abstract}

Using binary evolution with Case-C mass transfer, the spins of several black holes (BHs) in X$-$ray binaries (XBs) have been predicted and confirmed (three cases) by observations.  
The rotational energy of these BHs is sufficient to power up long gamma$-$ray bursts and hypernovae (GRBs/HNe) and still leave a Kerr BH behind.  
However, strong magnetic fields and/or dynamo effects in the interior of such stars deplete their cores from angular momentum preventing the formation of collapsars.  
Thus, even though binaries can produce Kerr BHs, most of their rotation is acquired from the stellar mantle, with a long delay between BH formation and spin up.  Such binaries would not form GRBs.
We study whether the conditions required to produce GRBs can be met by the progenitors of such BHs.  
Tidal-synchronization and Alfv\'en timescales are compared for magnetic fields of different intensities threading He stars. 
A search is made for a magnetic field range which allows tidal spin up all the way in to the stellar core but prevents its slow down during differential rotation phases.
The energetics for producing a strong magnetic field during core collapse, which may allow for a GRB central engine, are also estimated.
An observationally-reasonable choice of parameters is found ($B \lesssim10^2$ G threading a slowly-rotating He star) which allows Fe cores to retain substantial angular momentum.  Thus, the Case-C-mass-transfer binary channel is capable of explaining long GRBs.
However, the progenitors must have low initial spin and low internal magnetic field throughout their H-burning and He-burning phases.

\end{abstract}



\section{Introduction}
\label{sec-Intro}

\subsection{Long GRBs}\label{subsec-lGRBs}

Long Gamma-Ray Bursts (lGRBs) are thought to be produced by the core collapse of rapidly-rotating massive stars.
Given their association with broad-line Type-Ic supernovae \citep[SNe; see, e.g.,][for a review]{2012grbu.book..169H} their immediate progenitors likely have neither hydrogen nor helium in their envelopes.
Thus, they are adequately described by the Collapsar model \citep{1993ApJ...405..273W,1999ApJ...524..262M}.

In the Collapsar scenario the core of a Wolf-Rayet (WR) star collapses and forms a proto-neutron star (PNS). 
Due to the large mass of the stellar core, the shockwave created by the bounce of the PNS stalls and fails to produce a SN explosion.
Accretion of high-angular-momentum material turns the PNS into a Kerr black hole (BH).
This mechanism also rapidly forms an accretion disk that feeds the central compact object.  
Material along the rotational axis of the star has little angular momentum so it falls freely and produces a {\it lower} density region, clearing a path for the outgoing GRB jet.

Two significant sources of specific energy are available to a newly-formed stellar-mass BH during core collapse.
The first one is the gravitational binding energy of the infalling material.  
A Schwarzschild BH can have an efficiency of $\sim5.7\%$ in converting the rest mass of the infalling material into energy (from large radii down to the marginally stable orbit),
whereas a maximally rotating BH (i.e., $a_\star \equiv Jc/GM^2_{BH} = 1$, $J$ being the angular momentum of the BH, $M_{BH}$ its mass and $G$ and $c$ the gravitational constant and the speed of light, respectively) can reach $\sim 42\%$.
The second source is the rotational energy of the BH.
This can be as large as $\sim 29\%$ of the mass of a maximally rotating BH and can be extracted, e.g., via the Blandford-Znajek mechanism \citep[BZ;][]{1977MNRAS.179..433B}.

Both energy sources have one or two orders of magnitude more energy than required to explain the energies involved in gamma-ray bursts/hypernovae (GRBs/HNe).  
This allows for efficiencies to be at play during the lGRB/HN production.
Mechanisms tapping either of the two energy sources need rapidly rotating progenitors to allow for efficient energy conversion as well as accretion disk formation.
In this paper we will mainly refer to the BZ mechanism for the central engine powering the GRB/HN explosion given that in order to explain the longer GRBs, those lasting several-hundred to thousands of seconds, the neutrino-powered mechanism needs stars with masses on the order of $\sim 100 \msun$ (assuming accretion rates of order $\sim 0.1\msun$ sec$^{-1}$ in order to maintain the required neutrino flux; see, e.g., \citealp{2002ApJ...577..893L,2005ApJ...632..421L,1999ApJ...518..356P,1999A&A...344..573R}), 
whereas BZ can explain them with less-massive pre-collapse stars given it requires lower accretion rates \citep[see, e.g., discussions in][]{2009MNRAS.397.1153K,2010MNRAS.401.1644B,2011arXiv1105.4193W}.

Another important ingredient for the BZ mechanism is a strong magnetic field ($\gtrsim 10^{14}$ Gauss) in the central engine.
Whether these are fossil fields (i.e., present at zero-age-main-sequence, ZAMS) or are produced by a dynamo during stellar evolution, a common envelope (CE) phase, or core collapse is unknown.
In this letter we will make the argument that if binaries are to explain lGRBs through the Collapsar model, then the fields must be generated during stellar core collapse.

\subsection{Binaries}\label{subsec-Binary}

The requirement of rapidly-rotating stars as progenitors of lGRBs/HNe provides strong constraints on their evolution.
This is because most massive stars will lose a large fraction of their ZAMS mass and angular momentum during their lifetimes \citep[see, e.g.,][]{2005ApJ...626..350H}.
Hence, single massive stars usually require low metallicity to acount for little wind mass loss \citep[see, e.g.,][]{2001ApJ...550..372F}. 
Differential rotation during late evolutionary stages is also necessary. 
Thus, weak coupling between different nuclear burning shells is, most likely, needed, i.e., weak magnetic fields.
Alternatively, models with low metallicity and strong rotational mixing that allow the stars to evolve chemically homogeneously also seem to work well \citep[see, e.g.,][]{2005A&A...443..643Y,2006ApJ...637..914W}.  

There is, nonetheless, an alternative which allows the tapping of a large external reservoir of angular momentum.
There are strong indications that most massive stars are in binary (or multiple) systems \citep[$\sim 71\%$; see][]{2012Sci...337..444S}.
If a fraction of these systems were to transfer part of their orbital angular momentum into stellar-spin angular momentum then there would be more than enough to produce a Kerr BH (e.g., $a_\star \gtrsim 0.3$) with an accretion disk around it during the gravitational collapse.

There are a few evolutionary paths suggested to achieve Kerr-BH progenitors from binary evolution 
\citep[see, e.g.][]{1998ApJ...494L..45P,2000NewA....5..191B,2002ApJ...575..996L,2003ARep...47..386T,2004MNRAS.348.1215I,2004ApJ...607L..17P,2007ARep...51..308B}.
However, one has to make sure that the transfer of angular momentum occurs late enough in the evolution such that it is not lost again, e.g., with the mass loss during the He-core-burning stage.
Case-C mass transfer \citep[after core-He exhaustion in the primary;][]{1970A&A.....7..150L} has been suggested as a viable path \citep[see, e.g.,][]{2002ApJ...575..996L,2007ApJ...671L..41B,2011ApJ...727...29M}.
Case M, i.e., tidally-synchronized, rapidly-rotating, chemically-homogeneous massive stars (usually with $M\!>\!40~\msun$) in a short-orbital-period binary (and where the components do not fill their Roche lobes, RLs), is a possible alternative \citep[see][]{2008IAUS..252..365D,2009A&A...497..243D}.
An estimate of the Kerr parameter of the resulting BH(s) from Case-M evolution would be similiar to that of the Case-C scenario.
A late merger, possibly after Case C, is not out of the question as well \citep{2005ApJ...623..302F}; however, the merging companion must be H devoid so as to produce a Type Ib/c HN.

The assumptions involved in the Case-C mass transfer scenario are justified as it has already correctly predicted the Kerr parameters of three Galactic sources (4U 1543$-$47, GRO J1655$-$40 and XTE J1550$-$564) with nuclear-evolved companions (see the X-ray-continuum-fitting, as well as Fe K-line in one case, observations by \cite{2006ApJ...636L.113S} and \cite{2011MNRAS.tmp.1036S}).
Another extragalactic source, LMC X$-$3, also seems to be consistent \citep[see][]{2006ApJ...647..525D} with the predicted value if one assumes that part of the rotational energy went in to powering up a Cosmological GRB.
However, if one applies similar assumptions to the evolution of BHs in high-mass X-ray binaries (HMXBs) the predicted BH spin parameters are well below what the observations suggest.
Moreno M\'endez \& Cantiello (Submitted) show that these spins cannot be acquired during the collapse of the star, e.g. with a spiral SASI \citep[Standing Accretion Schock Instability;][]{2000A&A...363..174F,2002A&A...392..353F,2007ApJ...654.1006F,2010ApJ...725.1563F,
2011ApJ...732...57R,2013ApJ...770...66H} as in the case of neutron stars \citep{1998Natur.393..139S,2007Natur.445...58B}.
This situation is not unreconcilable if hypercritical (or super-Eddington) mass accretion from a wind Roche-lobe overflow \citep[wRLOF;][]{2007ASPC..372..397M} can spin up the BHs after formation as shown in \citet{2008ApJ...689L...9M} and \citet{2011MNRAS.413..183M}.
But it does issue a warning that some assumptions may have to be carefully tested to assess their validity range.
(On the other hand, if the spins of the BHs in the HMXBs were natal their available energies would be well above $10^{54}$ ergs!).

\citet{2012ApJ...752...32W} point out that, although close binaries with a He star might be able to produce Kerr BHs, they might not be able to produce lGRBs as the Fe core has been depleted of most of its angular momentum by transferring it to the He envelope (which is accreted much latter during the collapse).
In this paper, I want to point out that this is correct under the assumption that a  magnetic field, strong enough to power a GRB, threads the He star and strongly couples the core to the mantle.
Nevertheless, there are regimes in both tidal-synchronization spin up and magnetic-field intensity, which allow massive stellar cores to keep most of the angular momentum during the subsequent contraction stages.
Furthermore, the large magnetic field necessary for a BZ-like central engine can still be in place, in time, to power the lGRB/HN explosion.

In section~\ref{sec-Rotation} we discuss the angular momentum conditions necessary to produce a collapsar GRB.
In section~\ref{sec-Timescales} the relevant timescales to transfer and maintain angular momentum in the core of the He star (progenitor of the Kerr BH) are described.
In subsection~\ref{subsec-resynch} we first verify the very fast depletion of angular momentum from the cores as estimated by \citet{2012ApJ...752...32W} and, second, estimate the magnetic fields required to allow for the core to be spun up (by tides) but not slowed down (by Alfv\'en torques due to contraction).
Section~\ref{sec-CentralEngine} describes a model capable of producing a GRB central engine, and estimates the energy requirements to produce, during core collapse, the magnetic field necessary for the BZ model to extract the energy.
Section~\ref{sec-Conclusions} provides a discussion of implications of the results and shows the conclusions.

\section{On The Necessary Angular Momentum for Collapsars}
\label{sec-Rotation}

As mentioned above, the main requirement for the formation of a Collapsar-induced GRB is the formation of an accretion disk.  
This implies that part of the material falling towards the BH must have more angular momentum than that required to orbit the BH.
In general relativity, unlike Newtonian gravity, there is a limit to how small the orbital radius can be (other than the surface of the attractor).
For the Schwarzschild metric ($a_\star = 0$, i.e., no rotation) the event horizon of a BH is at $R_{Sch} = 2 GM_{BH}/c^2$ 
the marginally bound orbit is at twice this radius, $R_{mb} = 4 GM_{BH}/c^2$ and the marginally stable orbit is at $R_{Sch}$, i.e., $R_{ms} = 6 GM_{BH}/c^2$.  
However, for an extreme Kerr BH ($a_\star = 1$), $R_{Kerr} = R_{mb} = R_{ms} = 1 GM_{BH}/c^2$ \citep[where $R_{Kerr}$ is the radius for the event horizon of a Kerr BH; see, e.g.,][]{1973grav.book.....M}.
It is common to assume that the innermost stable circular orbit (ISCO) is at $R_{ISCO} = R_{ms}$, given that a particle at $R_{mb}$ has more angular momentum and energy. 
Nonetheless, during hypercritical accretion, this is not always necessary, and the ISCO can be pushed further in, to the marginally bound orbit.
It has also been shown by \citet{2006ApJ...641..961L} that material with enough angular momentum to orbit at $R_{mb}$ may produce dwarf disks which, in turn, may hold back material with less angular momentum (than that necessary to be at the ISCO) on their outer rim.
In any case, as the PNS turns into a BH (at around 3 $\msun$) the minimum specific angular momentum ($l_{ISCO}^{\,min} = L_{ISCO}^{\,min}/M_{BH}$, where $L$ is the angular momentum) required to form an accretion disk is around $l_{ISCO}^{\,min} \simeq 10^{16.7}$ cm$^2$ s$^{-1}$.

Fig.~2 of \citet{2007Ap&SS.311..177V} shows a plot of the minimum necessary specific angular momenta to form disks around $a_\star = 0$ and $a_\star = 1$ BHs (using $R_{ISCO}=R_{mb}$) as well as the specific angular momentum of equatorial material for an 8-$\msun$ He star synchronized with a 0.8-$\msun$ companion in a 7.2-hour orbital period.
In this scenario the companion star is filling its Roche lobe.
Thus, at this stage, all the equatorial He-star material would fall into a disk before falling into the BH.
However, as \citet{2012ApJ...752...32W} show (see their figs.~4 through 6), this angular momentum can be efficiently extracted from the core by the time the stellar core is about to collapse.
From their Fig.~5 we can even appreciate that all of their models keep enough angular momentum until He core depletion and some of their models keep enough angular momentum until carbon ignition.
Thus, as we will show in the next section, GRB progenitors in binaries must evolve with low magnetic fields.


\section{Timescales}
\label{sec-Timescales}

We concentrate mainly on binaries that survive BH formation.
This is not because mergers (or binaries unbound by mass lost due to the lGRB/HN event) cannot produce collapsars, but rather because we wish to show that the former can also explain them.
Furthermore, surviving binaries allow a reconstruction of the BH formation process. 
This is because the nonsymmetric mass loss (off the center of mass) in the binary is {\it recorded} by the Blauuw-Boersma kick \citep{1961BAN....15..265B,1961BAN....15..291B,2008ApJ...685.1063B}, the orbital period, the orbital eccentricity, the miss-alignment of the angular momenta of the binary and the two stars, the chemically enriched companion 
\citep[see, e.g., the case of GRO J1655-40 which was studied by][]{1999Natur.401..142I,2008A&A...478..203G}, etc.  
Instead, detecting a single BH is nearly impossible and studying its formation history is hopeless; unless we were to observe a nearby (Galactic) lGRB/HN.

\subsection{Case C Mass Transfer}\label{subsec-CaseC}

A crucial piece in the model of \citet{2002ApJ...575..996L} and \citet{2011ApJ...727...29M} allowing an estimate of the natal spins of the BHs is that the primary star must be tidally synchronized and with close to solid-body rotation soon after ignition of core-carbon burning.
Case-C mass transfer followed by a CE phase allows the binary to evolve into a close orbit at the expense of removing the H envelope of the primary.
The proximity of the binary components compared to the primary-star radius allows for rapid synchronization with the orbit and, thus, a large spin late in the evolution of the primary (see section~\ref{subsec-tides}).
This evolutionary path prevents the loss of too much mass early in the evolution and allows for the formation of a massive core that leads to the formation of a Kerr BH.

\citet{2011ApJ...727...29M} assumed that the star reaches the collapse stage while rotating as a solid body.
This was, however, an over simplification.
What was estimated at that point was the remaining total angular momentum that the He star had, and whether that was enough to produce a rapidly rotating BH.
As \citet{2012ApJ...752...32W} point out, the internal distribution of the angular momentum may play a crucial role on whether or not a GRB central engine may be produced.
In the scenario by \citet{2011ApJ...727...29M}, the total angular momentum of the star will remain nearly constant given the tidal synchronization with the companion and the small amount of material lost through winds in the remaining stellar lifetime (i.e., in the remaining $10^3$ to $10^4$ years even at $\dot{M} \simeq 10^{-4} \msun$ yr$^{-1}$ only $\Delta M\!\lesssim\!1 \msun$ is lost from a He star with $M_{He}\!\gtrsim\!10 \msun$).
In a sense this synchronization is a large reservoir of angular momentum which the star can tap until it collapses.
Nevertheless, angular momentum may still be drained from the core to the outermost helium layer if the timescales required to re-synchronize the rotation of different shells of the evolved star as they contract and spin up are too short.
We will estimate the conditions that allow the contracting shells to retain a substantial amount of their specific angular momentum in section~\ref{subsec-resynch}.

\subsection{Tidal-Synchronization Timescales}\label{subsec-tides} 

After the CE phase, the primary has lost its hydrogen envelope leaving a helium or carbon core exposed.  
The orbital separation has decreased from $\sim1,500 \;\rsun$ to a few solar radii. 
What brings the CE phase to an end is still under debate but what is certain is that, in many binaries, the primary (and maybe the secondary) fills a large fraction of its Roche lobe.
This allows for a  small $a/R$ ratio ($a$ the orbital separation and $R$ the stellar radius) and a short synchronization timescale.
From \citet{1975A&A....41..329Z} we know that the dynamical-tide-induced-synchronization timescale is 
\be
\tau_{DT} = \frac{2^{-5/3}E_2^{-1}}{5q^2(1+q)^{5/6}} \left(\frac{R^3}{GM}\right)^{1/2}\frac{I}{MR^2} \left(\frac{a}{R}\right)^{17/2}.
\ee
where $R$ is the stellar radius, $G$ is the gravitational constant, $M$ is the mass of the star, $q = M_2/M$ ($M_2$ the mass of the companion star), $I$ is the moment of inertia of the primary star, $a$ is the orbital separation and $E_2$ is a tidal-torque costant dependent on the stellar structure \citep[a table for these is available in][]{1975A&A....41..329Z}.
And, from \citet{1977A&A....57..383Z}, we have that the equilibrium-tide timescale is
\be 
\tau_{ET} = \frac{t_F}{6q^2\lambda_{sync}} \frac{I}{MR^2} \left(\frac{a}{R}\right)^6,
\ee
where $\lambda_{sync} \simeq 0.02$ \citep[see][]{1989A&A...220..112Z,2012ApJ...752...32W}.
Here $t_F$ is the friction timescale which measures the efficiency of the viscous dissipation and can be expressed by
\be
t_F = \frac{M_{env}R^2}{L}
\ee
with $L$ being the stellar luminosity and $M_{env}$ the mass of the stellar envelope.

Again, it is debated which timescale is more adequate for massive stars \citep[see, e.g., the discussion of][]{2007A&A...461.1057T}, especially when the envelope has been stripped away. 
Nonetheless both approximations seem to work on a timescale similar to the lifetime after He-shell burning for massive stars filling a large portion of their RLs, i.e., a few hundred to a few thousand years \citep[see, e.g.,][]{2007Ap&SS.311..177V}. 

For the predictions of \citet{2002ApJ...575..996L}, \citet{2007ApJ...671L..41B} and \citet{2011ApJ...727...29M} to work, as the observations by \citet{2006ApJ...636L.113S} and \citet{2011MNRAS.tmp.1036S} suggest, it is necessary that the primary star fully synchronizes 
after Case-C mass transfer.  
In a sense, these observations suggest that (a) tidal synchronization occurs, (b) the angular momentum acquired through this synchronization process is conserved before and during core collapse and (c) the GRB/HN explosion does not drain too much rotational energy from the Kerr BH.

\subsection{Alfv\'en Timescales}\label{subsec-resynch}

As the core of the He star contracts to burn carbon it further spins up, conserving angular momentum.  
The C-burning stage of the core only lasts a few hundred to a few thousand years.
Further down its evolution, the core will further contract (and spin up) as it burns neon, oxygen, magnesium and, eventually, silicon into iron \citep[see, e.g.,][]{1990RvMP...62..801B}.
Each of these stages are ever shorter and so the question of whether they will slow down and re-synchronize with the rest of the star (i.e., solid body rotation as opposed to differential rotation) becomes relevant.
 
As an example, let us briefly discuss the case of a He star with a spin period of 7 hours ($1.44\times10^4$ seconds)  which rotates as a solid body.
Before C ignition, the inner $1.5 \msun$ has a radius of $\sim 2\times\!10^{10}$ cm; when it becomes an Fe core its radius will be of the order of a few $\sim 10^8$ cm and its period will be of the order of a few seconds;
by the time the core collapses to a PNS, its radius will be a few times $10^6$ cm and its spin period will be of the order of tenths of a millisecond (i.e., specific angular momentum $j\!=\!J/M\!\simeq\!4\times10^{16}$ cm$^2$ s$^{-1}$).

We must now estimate the timescale over which the magnetic field torques synchronize the contracting core with the outer layers of the star.
The important timescale is that of the Alfv\'en waves, given that this is the timescale on which the ions of the plasma react. 
Thus, the Alfv\'en timescale will be the time on which the shells of the star react to the magnetic torques created by differential rotation due to contraction of the core.  
These can be estimated by \citep[see, e.g.,][]{1962clel.book.....J}:
\be
\tau_{\rm A}\!=\!\frac{R_c\sqrt{4\pi\rho}}{B},
\ee
where $R_c$ is the radius of the core, $\rho$ is its density and $B$ is the magnetic field threading the contracting core.
If we assume that the needed $B\!\simeq\!10^{15}$ G field (for the GRB central engine) is formed from magnetic flux conservation (notice this is a big assumption given that the core is convective) then we need to begin with a field $B\!\simeq\!10^5$ to $10^7$ G threading the He star (depending on whether the radius of the region where we need the $10^{15}$ G magnetic field is $10^6$ or $10^7$ cm).

\begin{table}
\begin{center}
\begin{tabular}{|c|c|c|c|c|}
\hline
      Layer     &       Mass      &      Radius      &           Density        &         B Field        \\
   Composition  &   $ [\msun]$    &   $[\rm cm]$     &    $[\rm g \;cm^{-3}]$   &         [Gauss]        \\
\hline  
\hline
        He      &        $5$      &     $10^{11}$    &   $\frac{3}{4\pi}10^1$   &     $10^5$ - $10^7$    \\
       C-O      &       $10$      &     $10^{10}$    &   $\frac{6}{4\pi}10^4$   &     $10^7$ - $10^9$    \\
        Fe      &      $1.5$      &       $10^8$     &   $\frac{9}{4\pi}10^9$   &  $10^{11}$ - $10^{13}$ \\
\hline
\end{tabular}
\end{center}
\caption{Toy model of a pre-core-collapse He star of $16.5 \msun$.  Loosely based on the mass ratios in Fig. 33.1 of \citet{1990sse..book.....K}.  The column for the magnetic field is based on the requirements to produce a $10^{15}$ G field in a region of $10^6$ to $10^7$ cm around the BH.}\label{Tab:ModelStar}  
\end{table}

Again, let us use a toy model of a $16.5 \msun$, pre-core-collapse, He star with the composition in table~\ref{Tab:ModelStar}.  
Note that, using the relation between the He core and the ZAMS mass of the star \citep[see, e.g.,][]{2002ApJ...575..996L}
\be
M_{He} = 0.08 \left(\frac{M_{ZAMS}}{\msun}\right)^{1.45}\msun
\ee
we know that such a star must have originally been around $40 \msun$. 
For this scenario we obtain the following minimum Alfv\'en timescales (starting withing $B\!\simeq\!10^7$ G; for a field that starts with $B\!\simeq\!10^5$ G, the timescales are a factor of $10^2$ longer):
\be
\tau_{\rm A,Fe} \simeq 1  \,{\rm sec},\,\,\, \tau_{\rm A,C} \simeq 2,450 \, {\rm sec} \,\,\, \nonumber
\\{\rm and} \,\,\, \tau_{\rm A,He} \simeq 5.5 \times 10^4  \,{\rm sec}.
\ee

To put this in perspective of late-stellar evolution, the Si core burns into an Fe core in about a day; the C core takes about a thousand years to burn.  
Of course, many nuclear-burning steps are missing in our estimate, but we can clearly see that the Alfv\'en timescales are orders of magnitude shorter than their corresponding nuclear-burning timescales.  
We can safely assume that in such a scenario stars will not have differential rotation but will rather rotate as rigid bodies.

Since
\[
\tau_A\!\propto\!\frac{R_c\cdot R_c^{-3/2}}{R_c^{-2}}\!\propto\!R_c^{3/2}
\]
the synchronization timescales become smaller as the core contracts into further burning stages.
In table~\ref{Tab:AlfvenTS} we can observe how the timescales grow longer as we relax the condition of forming the Magnetar-strength magnetic field from a fossil field.
For a pulsar-like field, the Alfv\'en timescale and the nuclear-burning timescales are similar for the iron core. The carbon burning timescale is somewhere between $\tau_{A,C}$ and $\tau_{A,He}$ for the case where the final magnetic field is $10^{10}$ G. 

\section{A Model For Generating A Central Engine}
\label{sec-CentralEngine}

To obtain a stellar core with enough angular momentum  a star needs to tidally synchronize and posses a magnetic field which allows angular momentum to be transferred on a timescale of the order of the nuclear timescale of carbon, i.e. $\tau_{sync} \simeq \tau_A \simeq \tau_{n,C}$.
These timescales must also prevent angular-momentum extraction from the core as further contraction and spin up occur.
A magnetic field of $B \!\lesssim\! 10$ G threading the He star  would likely produce the desired effect (from section~\ref{subsec-resynch}; through magnetic flux conservation, this would produce a magnetic field of $\sim 10^{9}$ G after core collapse).

\begin{table}
\begin{center}
\begin{tabular}{|c|c|c|c|}
\hline
Final Magnetic Field   &   $\tau_{\rm A,He}$   &   $\tau_{\rm A,C}$    &  $\tau_{\rm A,Fe}$  \\
    $ [\rm Gauss]$     &   $ [\rm Seconds]$    &   $[\rm Seconds]$     &   $[\rm Seconds]$   \\
\hline  
\hline
        $10^{15}$      &   $5.5 \times 10^4$   &   $2.5 \times 10^3$   &         $1$         \\
        $10^{12}$      &   $5.5 \times 10^7$   &   $2.5 \times 10^6$   &       $10^3$        \\
        $10^{10}$      &   $5.5 \times 10^9$   &   $2.5 \times 10^8$   &       $10^5$        \\
\hline
\end{tabular}
\end{center}
\caption{Alfv\'en timescales for the He, C and Fe shells estimated for fossil magnetic fields threading the star such that the magnetic field within $10^7$ cm around the compact object becomes that of the first column after core collapse.  The timescales grow by a factor of a 100 if the radius is reduced to $10^6$ cm.}\label{Tab:AlfvenTS}
\end{table}

Dynamos during core collapse should increase the magnetic field near the core by a factor of $10^6$ to allow for a BZ central engine to produce a GRB/HN.

As mentioned in section~\ref{subsec-lGRBs}, the available core-collapse energy from binding-energy release as well as from rotation can be quite substantial.  
Energies on the order of hundreds to thousands of Bethes have been inferred from observations of the rotation of several Galactic BH binaries \citep[see, e.g.,][]{2011ApJ...727...29M}.
It is easily shown that the energy necessary to create a magnetar-like field ($E_B$) around the PNS or BH will not require more than a diminute amount of this available energy\footnote{Notice our estimates are Newtonian, however the Relativistic corrections are small compared to the wide energy ranges we are considering.}.
\be
E_B = uV = \left(\frac{B^2}{8\pi}\right)\left(\frac{4\pi R_c^3}{3}\right)
\ee
where $E_B$ is the energy, $u$ is the energy density and $V$ is the collapsed-core volume of radius $R_c$ where the BZ mechanism will take place.  
For $B \simeq 10^{15}$ G, and $R_c \simeq 10^7$ cm (a few times larger than the volume of the PNS) we have
\be
E_B \simeq 10^{50} {\rm erg}.
\ee
For a more realistic $R_c \simeq 10^6$ cm (considering a Schwarzschild BH has an event-horizon radius $R_{BH} \simeq 1.5$ km $\msun$; that of a Kerr BH is even smaller and an accretion-disk-innermost-circular orbit, ISCO, is at a few times this radius, $R_{ISCO}\lesssim 3R_{Schw}$) the necessary energy drops to
\be
E_B \simeq 10^{47} {\rm erg}.
\ee
Hence, it is clear that producing such a field is energetically feasible.
In both cases $B$ is well below what an equipartition field could reach.  
Indeed, an equipartition magnetic field could become (although highly unlikely) as large as $10^{17}$ G for an available energy of $E \simeq 10^{54}$ erg.
Several recent studies show numerical calculations of core collapses where a SASI (Standing-Accretion-Shock Instability) or another mechanism produce a dynamo that leads to the formation of magnetic fields above $10^{14}$ G \citep[see, e.g., calculations and review in][]{2012ApJ...751...26E}.  Similarly, \citet{1993ApJ...408..194T} argue that during the cooling of the NS (or PNS, since convection stops when the NS becomes transparent to neutrinos) the field can be widely amplified. 

\section{Discussion and Conclusions}
\label{sec-Conclusions}

From the requirements shown in section~\ref{subsec-resynch} regarding the magnetic field threading through the core and the He layer, it is clear that stars which possess large magnetic fields, fossil or dynamo-produced, are not the best candidates to produce a long GRB during, or soon after, core collapse.

As mentioned in \citet{2012ARA&A..50..107L} and references therein, fossil fields are difficult to maintain during the protostar phase, nonetheless, dynamos can easily build a large field during the main sequence of the star.
Thus, eventually, preventing the formation of a rapidly rotating Fe core.
It is, therefore, highly desirable that the GRB-progenitor star is a slow rotator throughout its main sequence and He-burning stages.
This avoids the enhancement of the convective dynamos with differential-rotation dynamos, i.e., preventing an $\alpha-\Omega$ dynamo from being sustained \citep[e.g., see section 8 of][where they discuss high Rossby number dynamos, where $Ro \equiv P/\tau_{conv}$, with $P$ the rotation period and $\tau_{conv}$ the convective overturn timescale]{1993ApJ...408..194T}.

In this paper it has been shown that a massive star in a binary that undergoes Case-C mass transfer, a common envelope and tidal synchronization may retain most of the core angular momentum gained through tidal interactions if the magnetic field threading the stellar core and the He envelope is around $B_{He} \!\lesssim\! 10$ G.  
Furthermore, a magnetar-like field, $B\sim10^{15}$ G, may easily be produced during core collapse (e.g., via a SASI and/or convective dynamos) given the large amount of available energy during this phase of stellar evolution.

This scenario provides the necessary ingredients for the Blandford-Znajek mechanism to power GRB events, as the central object will be both, rapidly rotating and highly magnetized.  
Furthermore, the central region of the collapsing core will retain enough angular momentum to produce a rapidly rotating black hole and an accretion disk early after core collapse.

Thus, this model with the retention of angular momentum at the stellar core and the generation of strong magnetic fields during core collapse, facilitates the production of central engines capable of producing Cosmological GRBs and subluminous long GRBs, as well as the extremely energetic accompanying Type Ib/c hypernovae.

As previously suggested in, e.g., \citet{2011ApJ...727...29M}, there are at least 7 Galactic BH binaries which, from the estimates of their BH spins (and, in some of them, metal enrichment of the companion), are likely remnants of subluminous GRBs; they possess so much rotational energy that they likely dissmantle the central engine before the GRB has time to fully develop.  LMC X$-$3 is a good candidate of a progenitor of a Cosmological GRB \citep{2008ApJ...685.1063B}.  This binary, having a more massive companion, thus longer orbital period, has lower rotational energy, more suit up to allow the central engine to survive long enough to produce a long GRB.  Better estimates of the spin of the BH in this system, as well as system-velocity measurements of this binary system would provide valuable information to test the Case-C mass-transfer formation scenario for GRB/HN progenitors.

\section*{Acknowledgements}
\label{sec-Acknow}

The author had support from an Argelander Fellowship as well as a CONACyT fellowship and projects CB-2007/83254 and CB-2008/101958.  The author would like to thank William Lee, Dany Page,  Jonathan Braithwaite, Matteo Cantiello, Norbert Langer, Sergio Mendoza and Leonid Georgiev for useful comments, discussions and insight.  The author also appreciates the useful criticisms and comments from the anonymous referee.
This research has made use of NASA’s Astrophysics Data System as well as arXiv.



\end{document}